\documentclass[aps, onecolumn, notitlepage, floatfix,amsmath,nofootinbib, letterpaper,amssymb, showpacs]{revtex4-1} 
\usepackage{graphicx}
\usepackage{amssymb}
\usepackage{epstopdf}
\usepackage{slashed}
\usepackage{amsmath}

\usepackage{multirow} 

\newcommand{\ignore}[1]{}
\newcommand{\be}{\begin{equation}} \newcommand{\ee}{\end{equation}}
\newcommand{\ba}{\begin{eqnarray}} \newcommand{\ea}{\end{eqnarray}}
\newcommand{\nn}{\nonumber} 
\newcommand{\ra}{\rightarrow}

\renewcommand{\a}{\alpha} \renewcommand{\b}{\beta}

\newcommand{\p}{\partial}

\newcommand{\ran}{\rangle}
\newcommand{\lan}{\langle}

\def\slashb#1{\setbox0=\hbox{$#1$}#1\hskip-\wd0\dimen0=5pt\advance
        \dimen0 by-\ht0\advance\dimen0 by\dp0\lower0.5\dimen0\hbox
          to\wd0{\hss\sl/\/\hss}}
\input{epsf} 

\begin{document}


\title{Analysis Tools for Discovering  \\
Strong Parity Violation \\
 at Hadron Colliders}
 
 \author{Mihailo Backovi\'{c} and John P. Ralston } 
   \affiliation{Department of Physics \& Astronomy, \\ The University of Kansas,
  Lawrence, KS 66045}

 \begin{abstract}

Several arguments suggest parity violation may be observable in high energy strong interactions.  We introduce new analysis tools to describe the azimuthal dependence of multi-particle distributions, or ``azimuthal flow.'' Analysis uses the representations of the orthogonal group $O(2)$ and dihedral groups $D_{N}$ necessary to define parity completely in two dimensions.  Classification finds that collective angles used in event-by-event statistics represent inequivalent tensor observables that cannot generally be represented by a single ``reaction plane.''  Many new parity-violating observables exist that have never been measured, while many parity-conserving observables formerly lumped together are now distinguished.  We use the concept of "event shape sorting" to suggest separating right- and left-handed events, and we discuss the effects of transverse and longitudinal spin. The analysis tools are statistically robust, and can be applied equally to low or high multiplicity events at the Tevatron, $RHIC$ or $RHIC \, Spin$, and the $LHC$. 
 
 \end{abstract}

 \maketitle

\section{Many Forms of Strong Parity Violation}

{\it Why should strong interactions conserve parity?} Large violations of parity symmetry in strong interactions are possible by several independent means. Here we discuss new model-independent methods to probe strong parity violation at hadron colliders, including the Tevatron, $LHC$ and $RHIC$.  

\subsection{Strong Parity Violation at Low Energy}
Instanton approximations make one road to strong $CP$ violations \cite{'tHooft:1976fv}. The model is described by a term $ \theta \epsilon_{\mu \nu \a \b}tr(F^{\mu \nu}F^{\a \b})$ added to the QCD Lagrangian. Here $F^{\a \b}$ is the gluon field strength tensor, $tr$ indicates the trace over the colors, and $\theta$ is a parameter. Low-energy physics puts severe limits on the maximum size of such effects. Yet high energy collisions may well trigger a different vacuum phase with $CP$ violation of order unity.  Morely and Schmidt\cite{Morley:1983wr} long ago suggested testing strong $P$, $CP$, and $T$ violations in heavy ion collisions. Recently the topic has seen great activity\cite{Halperin:1998gx, Kharzeev:1998kz, Fugleberg:1998kk,Kharzeev:2004ey,Kharzeev:2007tn, Buckley:1999mv, Buckley:2000aa, Ahrensmeier:2000pg}, including significant experiments we will discuss.  Finch {\it et. al.}\cite{finch} discuss some of the history of this development. 

\subsection{High Energy Strong Parity Violation}

High energies offer new possibilities for strong parity violation. As such, they are 
natural targets of $RHIC$ and $LHC$ experiments. Consider for example a higher-derivative model given by the QCD Lagrangian plus $L_{6}$, where \ba L_{6} = {1 \over M^{2}} tr(\epsilon_{\mu \nu \a \b} F^{\mu \sigma }  F_{\sigma}^{\nu} F^{ \a \b} ). \label{newth} \ea In vector notation the $L_{6}$ Lagrangian goes like $c_{abc} \,  \vec E_{a} \times (\vec B_{b} \cdot \vec B_{c})$, where $\vec E_{a} $ and $\vec B_{a}$ are the color electric and magnetic fields, and $c_{abc}$ are color group structure constants The Lagrangian represents a chromo-magnetic effect with three to six gluon vertices. Its transformation properties are {\it odd } under time-reversal, {\it odd} under parity and  {\it odd} under $CP$. Very little is known to limit this Lagrangian. 

In contrast, 't Hooft's low-energy model goes like $\vec E_{a} \cdot \vec B_{a}$. It is a total 4-divergence naively dropping out of the integrated action, and giving ``zero'' observable effect on equations of motion. Its induced effects, expected from non-perturbative global subtleties of gauge fields, cannot really be estimated in perturbation theory.  In comparison, the effects of Eq. \ref{newth} are straightforward signals that naturally grow with energy. The effects amount to parity-violating multi-gluon vertices that have every reason to affect high energy multi-particle production. 

The coupling of $L_6 $ involves a mass scale  $M$, which new physics considerations would estimate in the $TeV$ range.  In the regime of small momentum transfers of $O(MeV)$, the ratio $(MeV/TeV)^{2} \sim 10^{-12}$ would greatly suppress its effects. It follows that constraints on the parameter $M$ from low energy phenomenology such as the neutronÕs electric dipole moment\cite{Weinberg:1989dx} are relatively weak for most beyond the Standard Model theories which can induce the operator in Eq. \ref{newth}.  Meanwhile the proposal of transient, local $P$ or $CP$-violating phase not probed by low-energy observables  may be even more natural when applied in a higher derivative context.  High energies are generally needed to probe higher derivative terms and any new phases they may produce. No further motivation for the $L_{6}$ model is needed other than an effective theory with gauge invariance and symmetry under the rest of the Lorentz group. 

Given that many other models of high energy strong parity violation exist, it is surprising that few if any have been tested. A common assumption that strong parity violating effects should be ``small'' has little experimental support at collider energies, except by extrapolation.  Testing strong $P$ and $CP$ symmetry at the new energy frontier  is now a priority.

\subsection{Focus on Signals}

While models have their purpose, it would be premature to concentrate on models when testing parity symmetry. The work here is concerned with developing {\it signals} of strong parity violation that do not depend on models. We focus on the symmetry properties of observables that might discover parity violation from any source. Recently the STAR\cite{star2010} experiment observed large effects in multi-pion azimuthal flow that are not consistent with certain parity-conserving simulations.  The effects are described as being ``a signal consistent with several of the theoretical expectations'' sensitive to charge separation predicted by strong $CP$ violation.  But what did STAR really measure? The experiment measured a {\it parity-conserving} statistic. 

We will review how a parity-even statistic became tied to concepts of ``odd parity fluctuations,'' sometimes leading to confusion over basic principles. We believe that reference to indirect, model-dependent deviations from a Monte Carlo simulation cannot possible compete with direct confrontation of data with observables that absolutely preserve or violate a symmetry.  As we will review in Section \ref{sec:manyangles}, the analysis long ago became embedded in an overly restrictive model - the ``standard event plane formalism.''  That step unknowingly made  parity-violating observables much harder to construct. It also greatly limited exploration of the full richness in multi-particle data with or without a focus on parity.  In the model-independent framework we present, the scope and power of event-by-event and Òazimuthal flowÓ statistics is not only retained, but  extended to many new observables, including a new concept we call ``shape sorting'' (Section \ref{sec:shapesort}).

Three-dimensional tests of parity symmetry tend to be complicated, and sometimes hinge on idealized three-dimensional detector symmetry.  Here we focus on two-dimensional tests because they are simple, and because the naturally high statistical power of multi-particle observables makes many {\it independent} tests possible. Which two dimensional projection one should use is in principle a matter of choice and experimental details. Due to recent interest in parity violation tests using azimuthal correlations, here we will focus on observables constructed in the transverse plane. 

It turns out that defining and detecting parity violation in two-dimensional azimuthal flow is quite subtle. First, we will show that {\it any quantity that is parity-odd by two dimensional tests is odd under ordinary three-dimensional parity}.  A proper two dimensional parity test procedure is {\it sufficient} to rule out parity symmetry, but not {\it necessary}: there also remains a number of independent three-dimensional tests.  

Any naive assumption that a single test would prove parity conservation is not true, due to deep geometrical features of the two-dimensional space.  ``Parity'' in two dimensions is a  rarely discussed subject that is more rich and complicated than the three dimensional rendition. By analyzing the transformation properties under $O(2)$, the orthogonal group in two dimensions, we not only discover new observable features of parity symmetry, but we also revise the {\it parity-conserving} measures of azimuthal flow.

\subsection{ Outline of the Paper} 

Besides testing parity with two dimensional quantities, the paper is concerned with the consistent two-dimensional statistical description of azimuthal flow. We find that inherent transformation properties cause significant revisions to standard formalism describing multi-particle data, whether or not parity symmetry is under discussion. 

The first step is group classification. The easy mapping of azimuthal moments into complex numbers produces distinct tensor transformation properties that have been Òhidden in plain sightÓ.  Transformation properties are crucial, because the distinction of different {\it tensors} is observable. Transformation under the dihedral group appears here, because {\it parity} in two dimensions needs it. 

The next step develops event-by-event statistics in general terms of conditional probabilities, also done in Section \ref{sec:azi}. It is very important not to impose models that bias the experimental description.  Section \ref{sec:parity} shows how to build variables capable of testing parity symmetry. Rather than singling out any ``best'' variable we show how to develop many {\it consistent} variables in order to let experimentalists determine what is ``best.'' This Section also explains the classification of the parity-motivated statistics used by STAR. Section \ref{sec:spin} discusses spin-dependent tests. 

\section{Many Reasons $2D$ Parity is So Interesting} 

Details of collider experiments and detector geometries often make it difficult to test for parity violation using  three-dimensional observables. It is thus more convenient to appeal to observables which ``live'' in two dimensional subspaces. In doing so, it is important to note that testing parity violation using two-dimensional observables should be consistent with three-dimensional parity violation.

In three dimensions parity is represented by {\it inversion}, $P_{3D}=-1_{3 \times 3}$. It is not necessary to invert all three axes to test parity symmetry, as we now show.

Consider any operation $D \in O(3)$ with determinant $det(D)= -1$. The operation of swapping $x$ and $y$ axes, with $z$ fixed, is an example that can loosely be called ``a parity transformation on a subspace.'' If symmetry of the Hamiltonian $H$ under $D$ fails, then $[ \, H,  \, D \,] \neq 0$. 

By construction one can always write $D=P_{3D}R$, where $R \in SO(3)$. When $D$-symmetry fails, then \ba [ \, H,  \, D \,] =[ \, H,  \, P_{3D}R \,] = P_{3D} [ \, H,  \, R \,] + [ \, H,  \, P_{3D} \,]R \neq 0, \nn \ea and then \ba [ \, H,  \, P_{3D} \,] \neq 0, \label{psym} \ea given rotational invariance $[ \, H,  \, R \,] =0$.  It follows that $P_{3D}$-symmetry is violated by finding {\it any single example} of parity violation on a subspace. 

Next: the map from three-dimensional data to two dimensions is not always trivial. Given a 3-momentum vector $\vec p =(p_{x}, \, p_{y},  \, p_{z})$, remove the $z$ component by projection: \ba (p_{x}, \, p_{y},  \, p_{z}) \ra (p_{x}, \, p_{y}). \nn \ea Apply the same projection to the 3-dimensional parity operator $P_{3D} =-1_{3 \times 3}$: \ba P_{3D}= \left(\begin{array}{ccc}- 1 & 0 & 0 \\0 & - 1 & 0 \\0 & 0 & - 1\end{array}\right) \ra  \left(\begin{array}{cc}-1 & 0 \\ 0 & -1\end{array}\right). \nn \ea The $2 \times 2$ operator has determinant +1: {\it it is a rotation}, not ``parity.''  The ordinary belief that parity is equivalent to an inversion fails! 

{\it Parity} is defined as the discrete subgroup with determinant $det=-1$ of the orthogonal group. We seek a matrix $P_{2D}$ which represents two-dimensional parity and is rotationally invariant.  Let $M_{inv}$ be a rotationally invariant matrix, \ba M_{inv} = \left(\begin{array}{cc} \a  & \b \\ \gamma & \delta \end{array}\right) . \nn \ea  Calculate the rotated matrix \ba R^{T}(\theta) \cdot M_{inv} \cdot R(\theta)= \left(\begin{array}{cc} cos\theta  & sin \theta \\ -sin \theta & cos \theta \end{array}\right) \cdot  \left(\begin{array}{cc} \a  & \b \\ \gamma & \delta \end{array}\right)\cdot \left(\begin{array}{cc} cos\theta  & -sin \theta \\ sin \theta & cos \theta \end{array}\right). \nn \ea Here $R(\theta) \in SO(2)$ and superscript $T$ stands for the transpose. Setting $R^{T}(\theta) \cdot M_{inv} \cdot  R(\theta) =M_{inv} $ gives $\delta =\a$, $\gamma = -\b$. All rotationally invariant matrices $M_{inv}$ then satisfy \ba M_{inv} &=&  \left(\begin{array}{cc} \a  & \b \\- \beta & \a\end{array}\right)  \:\:\:\:  \:\:\:\: (from \: rotational \: symmetry ).\label{rotsym} \ea Compute the determinant: \ba det(M_{inv}) &=& \a^{2}+\b^{2}>0. \nn \ea By requiring rotational symmetry, the condition $det(P_{2D}) =\a^{2}+ \b^{2} =-1$ is impossible. We find the remarkable fact that {\it no real rotationally-invariant operator exists for 2D-parity.} 

Candidates for 2D-parity must break rotational symmetry because all 2D matrices with $det=-1$ correspond geometrically to a mirror-reflection about some axis in the plane. For example $P_{0}= diag(1, \,-1) $ reflects $y$-components around the $x$-axis. Any rotation of the matrix will also have $det(R \cdot P_{0}\cdot R^{T})=det(P_{0})=-1$ and define a different ``parity.''  It follows that a continuous infinity of 2D-parity candidates exists: yet not one of them has the features expected in three dimensions.   

Despite these facts, it would be wrong to conclude that 2D parity does not exist as a consistent concept. 
The problem of ``2-parity'' is the lack of  a rotationally invariant {\it representation}. But in deriving Eq. \ref{psym} we used rotational symmetry of the {\it Hamiltonian}, and not the same symmetry of the parity {\it representation}. There are many cases in physics where an operator is not invariant, but {\it transforms} in a known way.  For example angular momentum is not a rotationally invariant quantity but we have rotationally invariant tests that it is conserved. For physical purposes of testing parity symmetry, we need invariant test criteria, whether or not the operator itself is invariant. That is, the {\it representation} of parity is a side issue compared to what we wish to test. 

Continuing: choose a candidate $P_{2D} \in O(2)$ by some axis convention. Let $\varepsilon$ be an operator to test $D$-parity symmetry; it will be odd under $P_{2D}$:   \ba P_{2D}^{T} \cdot \varepsilon \cdot  P_{2D} =-\varepsilon . \nn \ea Since $P_{2D}^{T}P_{2D}=1$ we find the trace $tr(\varepsilon)=0$. Impose rotational invariance of the test: $R^{T}\cdot  \varepsilon \cdot R=\varepsilon$. Rotational symmetry was solved by Eq. \ref{rotsym} for any invariant matrix $M_{inv}$.   $tr(\varepsilon)=0$ implies parameter $\a=0$.  Thus an invariant test exists. It is unique up to a scale $\beta$ we set to unity: \ba  \varepsilon = \left(\begin{array}{cc}0 & 1 \\ -1 & 0\end{array}\right). \nn \ea This is the 2-dimensional Levi-Civita symbol. It is the generator of $SO(2)$.  Since any generator is invariant under its own transformations, its rotational invariance is obvious. 
 
{\it Continuing}, the result did not depend on the choice of $P_{2D}$.  Therefore $\varepsilon$ is odd under every $det=-1$ transformation in $O(2)$, {\it regardless of the reflection axis.} Classifying the numerous representations is the topic of the next Section, using the dihedral group. The Section will show how our studies beginning with ``parity'' discover new features of azimuthal flow {\it with or without parity symmetry}.

\section{Azimuthal Flow Reconsidered}

\label{sec:azi}
 
Classification of transformation properties is fundamental to ``azimuthal flow.''  It is conventionally described as taking the Fourier transform of particle distributions in the azimuthal angle $\phi$ defined relative to the $z$ axis. By attending to transformation properties, we  find there is much more in properly describing azimuthal flow. Extra attention to the group representations of $O(2)$ and setting up event-by-event statistics pays off by revealing new parity violating observables.

\subsection{Dihedrals}

The group $O(2)$ has infinitely many distinct discrete subgroups, called the {\it dihedral groups} of order $N$, denoted $D_{N}$. 

Each $D_{N}$ group is developed from two generators ${\cal R}, \, {\cal P}$ with the defining properties \ba {\cal R}^{N}=1, \:\:\:\: {\cal P}^{2}=1, \:\:\:\: \, {\cal P}{\cal R}{\cal P}={\cal R}^{-1}. \nn \ea  The geometrical interpretation of ${\cal R}$ is a rotation by $2\pi/N$. The interpretation of ${\cal P}$ is a reflection about an unspecified fiducial axis in the plane. The property ${\cal P}{\cal R}{\cal P}={\cal R}^{-1}$ means the sense of rotations is reversed after the reflection. Although the generator algebra refers to no particular axis, it cannot be realized by matrices without selecting a preferred axis.

Each group $D_{N}$ has $2N$ elements $(R_{Nk}, \, P_{Nk})$, for $k=0...N-1$, with $det(R_{Nk})=1$, $det(P_{Nk})=-1$, corresponding to certain rotations and reflections. For $N=3$ an appropriate set is \ba R_{30}= \left(
\begin{array}{cc}
 1 & 0 \\
 0 & 1
\end{array}
\right),\:\:\:\:  R_{31}=\left(
\begin{array}{cc}
 -\frac{1}{2} & \frac{\sqrt{3}}{2} \\
 -\frac{\sqrt{3}}{2} & -\frac{1}{2}
\end{array}
\right), \:\:\:\:  R_{32}=\left(
\begin{array}{cc}
 -\frac{1}{2} & -\frac{\sqrt{3}}{2} \\
 \frac{\sqrt{3}}{2} & -\frac{1}{2}
\end{array}
\right); \nn \\ 
P_{30}= \left(
\begin{array}{cc}
 1 & 0 \\
 0 & -1
\end{array}
\right),\:\:\:\: P_{31}=\left(
\begin{array}{cc}
 -\frac{1}{2} & \frac{\sqrt{3}}{2} \\
 \frac{\sqrt{3}}{2} & \frac{1}{2}
\end{array}
\right),\:\:\:\: P_{32}=\left(
\begin{array}{cc}
 -\frac{1}{2} & -\frac{\sqrt{3}}{2} \\
 -\frac{\sqrt{3}}{2} & \frac{1}{2}
\end{array}
\right).
 \nn \ea 

Wigner\cite{wignerp591959} cites these matrices in his textbook as his first pedagogical example of an ``abstract group,'' and later develops it as a homework problem, with no mention of ``parity.''

The dihedral group is non-Abelian, except for the case of $N=1$, and group multiplication does not close until all elements are included. Because of that, one cannot make a consistent classification of two-dimensional ``parity'' by arbitrarily choosing one element. Nevertheless, by the previous relation to parity on 3-dimensions, any single violation of dihedral parity ($D-parity$) symmetry suffices to prove that 3-dimensional parity symmetry has been violated. As consistent, the invariant test $P_{Nk} \cdot \varepsilon \cdot P_{Nk} =-\epsilon$ holds for {\it all possible} values of $N, \, k$.

\subsection{Collective Angles}

Let $\Psi$ be a collective angle obtained from an event with $N$ particles in the final state. It is a random variable whose definition remains to be specified: it is {\it not the same concept} as the ``reaction plane.'' The joint distribution of any pion (say) at angle $\phi$ and $\Psi$ among a set of 3-momenta $p_{1}... p_{n}$ is defined by \ba f(\phi, \, \Psi) &=&   {d N \over d \phi  d\Psi} = \sum_{n} \,\int  \,  d^{3}p_{1}d^{3}p_{2}...d^{3}p_{n}\,  {d N \over d^{3}p_{1}  d^{3}p_{2}  ...d^{3}p_{n}      } \nn \\ & \times & \delta(\phi-\phi( \vec p_{1}, \, \vec p_{2}...\vec p_{n})) \, \delta(\Psi -\Psi( \vec p_{1}, \, \vec p_{2}...\vec p_{n}) ) . \nn \ea The expression leaves open the definition of $\Psi =\Psi( \vec p_{1}, \, \vec p_{2}...\vec p_{n}) )$ because there are many possibilities. The point is important, and developed below.  Supposing angles are constructed properly, any distribution with rotational symmetry will have a trivial marginal distribution $f(\Psi)$. Rotational symmetry then predicts a simple {\it conditional} distribution $f(\phi \, | \, \Psi)$,  \ba f(\phi \, | \, \Psi) = {  f(\phi \,,  \Psi) \over f(\Psi )}\ra  f(\phi -\Psi). \nn \ea  That implies the double-distribution does not need to be measured. It suffices to compute $\Psi$ event by event, and collect the $\phi-\Psi$ data into a single distribution. In effect events are ``rotated to add'' coherently.  

Such a distribution is conveniently expanded in a Fourier series: \ba f(\phi \, | \,  \Psi ) = \sum_{m} \, f_{m}e^{im \phi -im \Psi }. \nn \ea Reality of $f$ gives $f_{m}^{*}=f_{-m}$. This is an expansion in good $SO(2)$ ``angular momentum'' eigenstates, \ba  (-i { \p \over  \p \phi} )e^{im \phi} = m e^{im \phi}. \nn \ea 
Acting on $f_{m}$, the dihedral ($D_m$) transformations are\footnote{Note the transformation is anti-unitary, which is a convention inherited from the complexification of $f_{m}$. } \ba f_{m}  & \rightarrow &  { R_{mk}} f_{m}=e^{i 2\pi k/m}f_{m}; \nn \\ f_{m}   & \rightarrow &  { P_{mk}}  f_{m}=( e^{i 2\pi k /m}f_{m})^{*} . \nn \ea 

The angular momentum eigenstates are not eigenstates of dihedral parity, and vice-versa. That is consistent and necessary from operators that do not commute. Note index $m$ of $D_{m}$ is matched to the Fourier mode. Conversely, one can expand in a basis of good $D$-parity eigenstates: \ba  f(\phi  \, | \, \Psi ) = \sum_{m} \,  v_{m}cos \, m(\phi - \Psi )  + a_{m}sin \, m(\phi - \Psi) . \nn \ea  Then parameters $v_{m}$ are {\it even} and $a_{m}$ are {\it odd} under under all elements $P_{mk}$. With the understanding that the absolute value of angular momentum $J$ is implied, we indicate good $D$-parity quantum numbers by $|J|^{P}=J^{P}$.

There is an instructive but {\it wrong} way to over-simplify the classification.  An even (odd) transformation property of $cosine$  ($sine$) functions follows from the simplistic transformation $\phi \ra -\phi$.  Yet that operation, {\it which selects one particular origin of $\phi =0$,} is highly coordinate-dependent. Indeed the whole distinction between $sin \phi$ and $cos \phi$ hinges on a coordinate origin for $\phi=0$, which is arbitrary.  Basing decisions about ``parity'' on sine-versus cosine origin conventions cannot be physically meaningful. The over-simplification implicitly works under the assumption that only one mirror-plane about a reference point $\phi=0$ defines ``parity.''

It is somewhat more meaningful to examine the rotationally invariant functions of $\phi-\Psi$. We have seen that $D$-parity involves reflection through a reference axis in the plane.  Given a single pion at angle $\phi$, and an event with a single collective angle $\Psi$, it is natural to reflect through an axis defined by one or the other. But a single-reflection test can only confirm one (1) particular $D$-parity symmetry.  All the other $D$-parity transformations not even examined might contradict it. For example reflection symmetry of the distribution across an axis $90^{o}$ to $\Psi$ might well be violated even if reflecting across $\Psi$ itself was a good symmetry.  

Thus more work is needed.  Let $\phi_{*}$ be a reference point on the unit circle. Reflection of $\phi$ and $\Psi$ about the corresponding axis defined by $\phi^*$ is \ba \phi \ra \phi ' = \phi+ 2( \phi_{*} - \phi) = 2 \phi_{*} -\phi; \nn \\ \Psi \ra \Psi ' =\Psi  + 2(\phi _{*} -\Psi ) = 2 \phi_{*} -\Psi ; \nn \\ \phi-\Psi  \ra  \phi'-\Psi' =-( \phi-\Psi). \nn \ea  

Now $sin( \phi- \Psi)$ is correctly odd under all such reflections, regardless of axis orientations. That means it must be related to the operator $\varepsilon$ we found was unique.  The relation is very simple. Define multiplets \ba \hat x_{\phi} &=& (cos \phi, \, sin \phi); \nn \\ \hat x_{\Psi} &=& (cos \Psi, \, sin \Psi); \nn \\ then \:\:\:\: \:\:\:\: \:\:\:\: \hat x_{\Psi} \cdot \varepsilon \cdot \hat x_{\phi} &=&  sin(\phi - \Psi). \nn \ea As we discussed earlier, under $P_{Nk}$ this transforms like  \ba x_{\Psi} \cdot \varepsilon \cdot \hat x_{\phi}  \ra x_{\Psi} \cdot P_{Nk}^{T}\cdot \varepsilon \cdot P_{Nk} \cdot \hat x_{\phi} = -x_{\Psi} \cdot \varepsilon \cdot \hat x_{\phi} . \nn \ea Once more the choice of $P_{Nk}$ reflection does not matter.

\subsection{Many Distinct Collective Angles} 
\label{sec:manyangles}

Event-by-event analysis is a breakthrough \cite{voloshin98} whose repercussions are still being developed.  Our work focuses on transformation properties, and does not depend on the details of how the event sample is selected. We assume there are cuts on the magnitude of transverse momentum, rapidity, particle species, etc. without discussion.  We turn to reconsidering collective flow angles, so far represented by a single symbol ``$\Psi.$''

The arithmetic mean angle $\langle \phi \rangle$ is coordinate dependent, and transforms poorly.  Standard procedures of ``circular statistics'' \cite{batschelet} map data into $SO(2)$ representations using \ba   const \,( cos  \, m  \Psi_{ \, m \, }, \, sin  \, m \Psi_{m}) = \sum_{i} \, 
 ( cos  \, m \,  \phi_{i}, \, sin  \, m \,  \phi_{i}). \label{reps} \ea Here $\phi_{i}$ are angles of a data sample.  In event-by-event analysis, the index $i$ runs over a set of particle labels selected from each event, producing a number $\Psi_{m}$ from each event. 
 
By construction each $\Psi_{m}$ parameterizes an $SO(2)$ tensor of rank-$m$.  As a consequence, each multiplet transforms like $m$ units of $SO(2)$ angular momentum.  Upon rotating the $\phi$-coordinate system by angle $\delta$, this means 
 
 \ba  \left(\begin{array}{c} cos  \, m  \Psi_{ \, m \, }, \\ sin  \, m  \Psi_{ \, m \, }\end{array}\right) \ra \left(\begin{array}{c} cos  \, m  \Psi'_{ \, m \, }\\ sin  \, m  \Psi'_{ \, m \, }\end{array}\right)= \left(\begin{array}{cc}  cos m \delta  & -sin  m \delta  \\ sin  m \delta &  cos m \delta  \end{array}\right) \left(\begin{array}{c} cos  \, m  \Psi_{ \, m \, }\\ sin  \, m  \Psi_{ \, m \, } \end{array}\right). \nn \ea 

Since they transform differently each collective angle $\Psi_{m}$ is independent, and each represents new and distinct information about the underlying system. It is not correct mathematics to propose $\Psi_{1} =\Psi_{2} = \Psi_{3}...$ as a ``symmetry'' of a physical system. 

Similarly, in three dimensions different tensors of rank 1, 2, 3... cannot be identified as equivalent. Such tensors will be found event by event by multiplying momentum components $p_{j} p_{k} p_{l}...$, taking moments, and making conditional distributions. In no way would we imagine that the tensor of rank-3 (say) was kinematically determined by the tensor of rank-2 or rank-4.   

Closely related, and coming from Eq. \ref{reps}, each different $\Psi_{m}$ has the defining property \ba \Psi_{m} \equiv \Psi_{m}+ { 2\pi \over m}. \nn \ea The symbol $\equiv$ here means ``identical by definition.'' The angle $\Psi_{2}$ labels an axis, which transforms like a ``stick'' without oriented ends; by definition $\Psi_{2}=\Psi_{2}+\pi$. An angle identical to itself in the class $m=3$ does not represent anything geometrically like a vector, but something like the orientation of an ideal Mercedes-Benz symbol.  

The nearest equivalent to such properties in three dimensions comes from eigenvectors $\hat e_{\a}$ of real symmetric matrices, which have no sign.  A pitfall of ordinary notation allows one to write expressions such as $\hat e_{\a}\cdot \vec v$ and believe it means something invariant.  Each number $\hat e_{\a}\cdot \vec v$ is a scalar, yet ill-defined because $\hat e_{\a}\cdot \vec v \equiv -\hat e_{\a}\cdot \vec v$ when $\hat e_{\a}\equiv -\hat e_{\a}$. The sign issue seldom causes problems, because it cancels out in expansions of the form $\vec v = \sum_{\a} \, \hat e_{\a} \, (\hat e_{\a}\cdot \vec v)$.

Needless to say, one is not going to find the sign of an eigenvector by doing a physics experiment. The lesson is that conventional notation may fail to incorporate the transformation properties symbols have from their definitions. We cannot find a record of this being noticed for collective angles. There are many consequences. In our notation symbol $\Psi_{2}$ is the angle for {\it elliptic flow}. Consider the expressions ``$sin( \phi-\Psi_{2})$'', ``$cos( \phi-\Psi_{2})$'', which appear to be well defined. They appear to be rotationally invariant under the shift of origin $\phi \ra \phi+\delta, \, \Psi_{2} \ra \Psi_{2}+\delta$. Yet neither exists as a physical observable, because \ba sin( \phi-\Psi_{2}) \equiv sin( \phi-\Psi_{2}-\pi) =-sin( \phi-\Psi_{2} ); \nn \\ cos( \phi-\Psi_{2}) \equiv cos( \phi-\Psi_{2}-\pi) =-cos( \phi-\Psi_{2} ).  \nn \ea It follows that an experiment seeking these moments will either get zero, or obtain a bias created by an improper usage of $\Psi_{2}$.  

It may seem a paradox that related quantities such as ``$sin^{2}( \phi-\Psi_{2})$'' can be a good observable. Use the identity \ba sin^{2}( \phi-\Psi_{2}) ={ 1- cos(2 \phi - 2 \Psi_{2}) \over 2}. \label{indet} \ea This is manifestly a function of $2 \Psi_{2}$, and $2 \Psi_{2} \equiv 2\Psi_{2}+ 2 \pi$ is a safe periodic variable. Its relation to a second-rank tensor nature comes using index notation:  \ba sin^{2}( \phi-\Psi_{2}) = (\hat x_{\Psi} \cdot \varepsilon \cdot \hat x_{\phi} )^{2}=
 \hat x_{\Psi}^{i} \hat x_{\phi}^{j} \hat x_{\Psi} ^{k}\hat x_{\phi}^{l} \varepsilon ^{ij}\varepsilon ^{kl}. \nn \ea The product of four $\hat x$ makes a tensor ranging up to rank-4 which has been projected into a true scalar.  
 
These facts clarify puzzles coming from the {\it standard event plane formalism} \cite{Ollitrault:1992bk,Ollitrault:1993ba,Borghini:2001vi,Borghini:2001zr}. It is a {\it semi-classical model} in which a single unobservable event plane angle $\Psi_{RP}$ is used for all $\Psi_{m}$: 
\ba f_{RP}(\phi)  \equiv  \sum_{m} \, v_{m}cos(\phi-\Psi_{RP})+a_{m}sin(\phi-\Psi_{RP}). \label{RPform} \ea  Here symbol $\equiv$ mean a defining fact of the model. The model can be implemented with classical simulations in which reaction planes and multi-particle collisions are generated numerically. The authors of Ref. \cite{voloshin98, Voloshin:2008dg} recommend interpreting the different $\Psi_{m}$ as ``estimators'' of $\Psi_{RP}$, with discrepancies to be explained by fluctuations. {\it That step defines their model.} For model-independent analysis we recommend just the opposite, and exploring each $\Psi_{m}$ as the independent tensor variable it is. Whether or not concerned with parity symmetry, eachangular tensor variable represented by $\Psi_{m}$ought to be of physical interest, because each represents independent physical information. 

\subsection{Event-Shape Sorting} 
\label{sec:shapesort}

If one believed all collective angles $\Psi_{m}$ were equivalent, then all event angular shapes would be boring, and more or less summarized by ``elliptic flow.'' Let us shift emphasis to the {\it correlations} found in multi-particle distributions. Correlations in the distribution of different collective angles describe meaningful ``event shapes.''

Here is a simple example. An event's ``splash pattern'' can be characterized by first choosing a reference angle such as $\Psi_{2}$, the orientation of elliptic flow. By construction it is $D$-parity-symmetric relative to itself.  Each subsequent $m^{th}$ moment (a Fourier shape) has a particular offset relative to the second (elliptic) shape given by $\Psi_{m} -  \Psi_{2}$. The magnitude $a_{m}^{(2)}$ is a measure of the parity-odd strength of each particular mode. The sign is meaningful, and represents a sense of twist of $\Psi_{m}$ relative to $\Psi_{2}$. 

\begin{figure}[hbt]
\begin{center}
\includegraphics[width=3in]{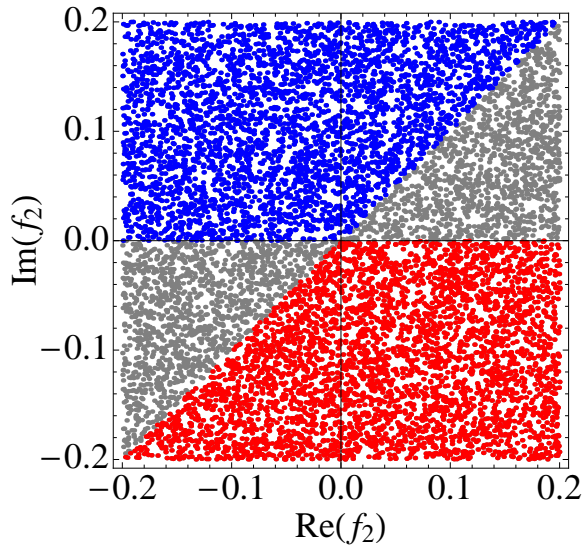}
\caption{ \small Real and imaginary parts of second Fourier moments sorted into Type-Minus (red online) and Type- Plus(blue online): see the text.  Third moments were fixed. }
\label{fig:SelectFou.eps}
\end{center}
\end{figure}

Rotational symmetry constrains the physics much less than one might believe. A general 3-variable distribution takes the form \ba  f( \phi, \, \Psi_{p},  \, \Psi_{q}) =\sum_{m} \, f_{m, \, m_{1},  \, m_{2}} \delta( m+ pm_{1} + q m_{2}) e^{i m \phi + i p m_{1}\Psi_{p}+  i q m_{2}\Psi_{q}}  \nn \ea This is more restrictive and more interesting than the naive periodic $f=f(\phi-\Psi_{1}, \, \phi-\Psi_{2})$.  It also depends on more variables than the model of Eq. \ref{RPform} is capable of emulating in principle. Thus for both parity-violating and parity-conserving circumstances, there will be advantages to using a completely model-independent formalism. 

 \begin{figure}[htb]
\begin{center}
\includegraphics[width=3in]{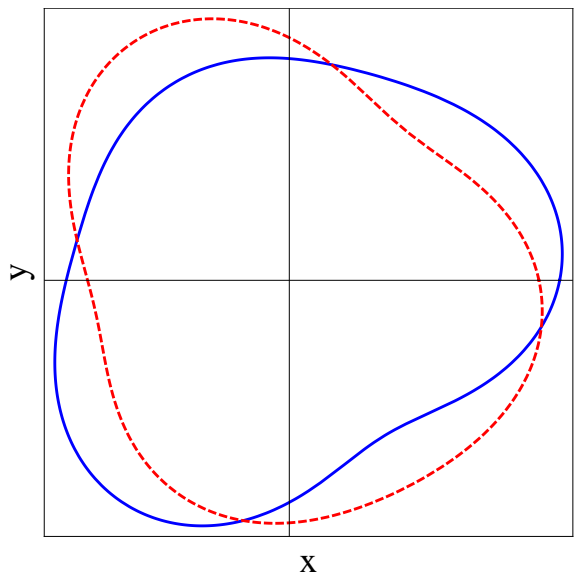}
\caption{ \small ``Event-shape sorting.''  Statistics of opposite $D$-parity are used to sort azimuthal angle data into classes labeled Type-Plus and Type-Minus. Sorted events make two distinct distributions, shown in polar plot as solid and dashed, where the radius at each angle equals the distribution. Data comes from Fig. \ref{fig:SelectFou.eps} as described in the text. }
\label{fig:SelectDist.eps}
\end{center}
\end{figure}

To explore such possibilities, we propose systematically recording event-by-event statistics to make distributions conditional on more than one collective angle, which we call ``event-shape sorting.'' We anticipate that strong interactions will develop interesting event shapes that grossly violate ordinary expectations. In particular, almost all individual events must violate $D$-parity. 

This may be surprising, because a naive interpretation of symmetry in quantum mechanics would expect that every single event be symmetric. However that is not observed.  Instead, parity and rotational symmetry in quantum mechanics are generally realized by events that actually break the macroscopic symmetry badly. Events generally come in right- and left-handed versions. Parity symmetry, if true, predicts equal numbers of left- and right-handed cases. After sorting, the physics of azimuthal flow makes particular shapes. Exposing different cases for study requires sorting the data into parity-determined types before making distributions.

\subsubsection{Simulation Example}

A Monte Carlo simulation illustrates the procedure.  A large number (5000) of 4-mode event distributions (modes $m=0...3$) was generated with flat random Fourier coefficients in the range keeping the distribution positive.  Distributions were generated so that the third moments $\hat x_{\Psi_{3}}$ were fixed.  The second moments were then sorted into two classes: Type-Plus with $Im(f_{2})<0$ and  $Im(f_{2})< Re(f_{2})$, (3787 events) and Type-Minus  (3826 events) with $Im(f_{2})>0 $ and $Im(f_{2})>Re(f_{2})$.  Fig. \ref{fig:SelectFou.eps} shows a scatter plot of the Type-Plus and Type-Minus moments. This particular selection merely illustrates one possibility. Fig. \ref{fig:SelectDist.eps} shows visually the Type-Plus and Type-Minus distributions accumulated over all events using a polar plot, and normalized by the number of events. The shape information represents average ``splash patterns'' coming out of the correlations in the events. Visual inspection shows the opposite $D$-parity. The (unsorted) sum of these two distributions is symmetric about the $x$-axis: the sum would pass statistical tests of $D$-parity symmetry based on $sin(\phi-\Psi_{1}) \ra -sin(\phi-\Psi_{1})$.  

Given the enormous multi-particle statistics of both the LHC and RHIC, we believe that many different combinations of event-shape observables will be statistically robust, and extremely interesting probes of strong interaction dynamics.

\begin{table}[htbp]
\begin{center} 
\resizebox {.95\textwidth }{!}{ 
\begin{tabular}{| c |c | c | c | c | c |} 
\hline Item & Product & Outcome &  Cartesian-Form &  Trig-Form & Parity \\  [0.1in]
\hline \hline
 
   &    &   $ 0_{AB}^+  $  & $\hat{x}_A \cdot \hat{x}_B$ &     $ cos(\phi_A - \phi_B) $ & +  \\ [0.05in]
1 & $1_A^\pm \otimes 1_B^\pm $    &   $0_{AB}^-$      &$\epsilon^{i \,j} x^i_A x^i_B$&   $ sin(\phi_A - \phi_B) $  & -    \\ [0.05in]
   &   &  $2_{AB}^+$    &$1/2( \hat{x}^i_A \hat{x}^j_B +  \hat{x}^j_A \hat{x}^i_B) - 1/2 \hat{x}_A \cdot \hat{x}_B $ &   
$
\mathcal{T}_{AB} =\left(
\begin{array}{cc}
  cos(\phi_A + \phi_B) &  sin(\phi_A + \phi_B)  \\
 sin(\phi_A + \phi_B)  &    -cos(\phi_A + \phi_B)\\   
\end{array}
\right)
$
 & +   \\ [0.25in]
\hline
    
    &    &   $0_{As}^+$      &$\varepsilon^{i \,j} x^i_A \hat{s}^j$&   $ sin(\phi_A - \phi_s) $  & +     \\ [0.1in]
  2  & $1_A^\pm \otimes 1_s^\mp $      &   $ 0_{As}^-  $  & $\hat{x}_A \cdot \hat{s}$ &     $ cos(\phi_A - \phi_s) $  & -  \\ [0.05in]
    &    &  $2_{As}^-$    &$1/2( \hat{x}^i_A \hat{s}^j +  \hat{x}^j_A\hat{s}^i) - 1/2 \hat{x}_A \cdot \hat{s}$&     $ \hdots $ & -   \\ [0.1in]
\hline
   
  &&&&&\\
  3 &   $0_{AB}^\pm \otimes 0_{AB}^\pm  $ &   $0_{AB \, AB}^+$      & $(\hat{x}_A \cdot \hat{x}_B)^2$ \,\, , \,\, $(\hat{x}_A \cdot \varepsilon \cdot \hat{x}_B)^2$ &   $ cos(2 (\phi_A - \phi_B)) $  & +  \\
  &&&&& \\
 \hline
   
  &&&&&\\
  4 &   $0_{AB}^\pm \otimes 0_{AB}^\mp  $ &   $0_{AB \, AB}^-$      & $(\hat{x}_A \cdot \hat{x}_B)(\hat{x}_A \cdot \varepsilon \cdot \hat{x}_B)$&   $ sin(2 (\phi_A - \phi_B)) $  & -  \\
  &&&&& \\
 \hline

   &     &   $0_{ABCD}^+$      &$tr(\mathcal{T}_{AB} \cdot \mathcal{T}_{CD})$&   $ cos(\phi_A + \phi_B - \phi_C - \phi_D) $  & +  \\[0.05in]
 5 & $2_{AB}^\pm \otimes 2_{CD}^\pm $   & $\vdots $   & $\vdots$          &    $\vdots$    &  \\ [0.05in] 
   &     &                   &                                   &    &  \\
 \hline
  
 &     &   $0_{AB \, \Psi_{RP}}^+$      &$tr(\mathcal{T}_{AB} \cdot \mathcal{T}_{\Psi_{RP} })$&   $ cos(\phi_A + \phi_B - 2\Psi_{RP}) $  & +  \\[0.05in]
 6  & $2_{AB}^+ \otimes 2_{\Psi_{RP}}^+ $   & $\vdots $   & $\vdots$          &    $\vdots$    &  \\ [0.05in] 
   &     &                   &                                   &    &  \\
 \hline
    
  &      &   $0_{AB \, C s}^-$      &$tr(\mathcal{T}_{A \,B} \cdot \mathcal{T}_{C \,s})$&   $ cos(\phi_A + \phi_B - \phi_C - \phi_s) $  & -     \\ [0.05in]
  7 & $2_{AB}^+ \otimes 2_{C s}^- $      & $\vdots $   & $\vdots$                                  &    $\vdots$          &            \\ [0.05in]
   &      &  & & & \\ \hline \end{tabular}  }
   
   \caption{ \small A survey of parity-even and -odd observables in two dimensions. ${\cal T}_{A\, B}$, etc. are the traceless symmetric parts of the matrices $\hat{x}_A^i \hat{x}_B^j$. The Item number identifies terms highlighted in the text.  Combinations not listed are indicated by dots.}
\label{tab:Table1}

\end{center}
\end{table}

\section{Testing Parity Symmetry}  
\label{sec:parity}

We proceed to construct parity symmetry-testing observables systematically.  Our focus is on general, model independent observables.

The ordinary Fourier moments $a_{m}$, defined relative to $\Psi_{m}$ case by case, transform as odd-quantities under $D_{m}$-parity. The most simple such observable is the diagonal $sine$ moment. However it vanishes {\it event by event} by definition: \ba  \sum_{i} \, sin  \, m \,  (\phi_{i}-\Psi_{m}) =   cos ( \, m \, \Psi_{m} ) \sum_{i} \, sin ( \, m \,  \phi_{i} ) -   sin ( \, m \, \Psi_{m}) \sum_{i} \, cos  (\, m \,  \phi_{i} )=0.\label{cancel} \ea  Each Fourier mode-$m$ already has $m$-fold rotational symmetry, as an eigenvector of $R_{mk}$. Each is effectively rotated into cosine form by the rule defining $\Psi_{m}$.
 
The physics lies in the correlations between different modes.  The ``best'' probe will be an experimental question.  For definiteness consider a model distribution $f=f(2\phi- 2\Psi_{2}). $ Let $a_{m}^{(2)}$, $v_{m}^{(2)}$ be the moments of the angular distribution relative to $\Psi_{2}$. The distribution is estimated by averaging over all data $\phi_{iJ}$ of all events $J$, numbering $N_{tot}$. A short calculation gives  \ba  {1 \over N_{tot}} \sum_{iJ} \, sin  \, m \,  (\phi_{iJ}-\Psi_{2}) ={1\over 2} a_{m}^{(2)}\, sin(\, m \, (2\Psi_{m} -  2\Psi_{2})); \nn \\ {1 \over N_{tot}} \sum_{iJ} \, cos \, m \,  (\phi_{iJ}-\Psi_{2}) ={1\over 2} v_{m}^{(2)}\, cos(\, m \, (2\Psi_{m} -  2\Psi_{2})). \label{fourierave}\ea Parity violation with a $\Psi_{2}$ angular reference probe is signaled by $a_{m}^{(2)} \neq 0$ 

Table \ref{tab:Table1} shows the classification of several quantities, which we now discuss. Recall our notation $J^{P}$ indicates the $SO(2)$ angular momentum absolute value $|J|$, and $D$-Parity $P$, simply called ``parity'' here.  The first moment $\hat x_{A}=(cos \phi_{A}, \, sin \phi_{B})$ transforms like $1^{\pm}$.  Classify the direct products $\hat x_{A}\otimes \hat x_{B}$.  A short calculation shows they decompose under $SO(2)$ into \ba 1^{\pm}\otimes  1^{\pm} \ra 2^{+}+0^{+}+0^{-}. \nn \label{groupreps} \ea Notice this is quite different from the $SO(3)$ rule $1\otimes 1 = 2 + 1 +0$. In more detail:   \begin{itemize}  \item $ \hat x_{A} \cdot \hat x_{B}  =cos(\phi_{A}-\phi_{B})$ is the only even-parity invariant ($0^{+}$) made from $\hat x_{A}  \otimes \hat x_{B} $.  When $\hat x_{A} \ra \hat x_{\phi}$ and $\hat x_{B} \ra \hat x_{\Psi_{1}}$ it has long been used to characterize directive flow: Item 1, Table 1. 
 
 \item $ \hat x_{A} \cdot \varepsilon \cdot \hat x_{B}  =sin(\phi_{A}-\phi_{B})$ is the only odd-parity,   $0^{-}$ available from $1^{\pm}_{A} \otimes 1^{\pm}_{B}$.  When $\phi_{A}, \, \phi_{B}$ are the azimuthal angles of pions with the $+$ and $-$ charges, it is the odd-parity observable of Kharzeev and Pisarski\cite{kahrz} liberated from reference to a $z$-axis. (See Table 1, Item 2 and Section \ref{sec:star}.)
 
\item A spin 4-vector $s^{\mu}$ of a Fermion projects into a $1^{ \mp}$ multiplet $\hat s$ in two dimensions with $D$-parity intrinsically opposite to $\hat x$. Then (Item 2, Table 1) $\hat x \cdot \hat s \sim 0^{-}$, and $\hat x \cdot \varepsilon \cdot \hat s  \sim 0^{+}$. See Section \ref{sec:spin} for more about transverse spin.   
 
 \item An interesting invariant (Item 6, Table 1) is made by reducing two spin-2 tensors to a new $0^{+}$ invariant. For definiteness, make the $J=2$ element from pions $A$, $B$, using the traceless symmetric tensor \ba {\cal T}_{AB, \, ij} = {\hat x_{i}(A) \hat x_{j}(B) +\hat x_{j}(A) \hat x_{i}(B) \over 2 } -\hat x(A) \cdot \hat x(B) \, \delta_{ij} . \nn \ea  Construct a similar spin-2 tensor ${\cal T}_{\Psi}$ from the collective flow multiplet $\hat x_{\Psi}$. The angle $\Psi_{2}$ describes the orientation of the tensor's principal eigenvector, $mod(\pi)$. The unique $0^{+}$ from combining the two tensors uses the trace: \ba 2_{AB}^{+} \otimes 2_{\Psi}^{+} \ra 0_{AB \Psi}^{+} &=& tr( {\cal T}_{AB} \cdot {\cal T}_{ \Psi}  ), \nn \\  &=& cos( \phi_{A} + \phi_{B} -2 \Psi). \label{volsoshinv} \ea

\end{itemize}

One important rule of thumb is that {\it functions of $ \, m_{1}\phi-m_{2}\Psi_{m_{p}}$ only exist as invariant observables when $m_{1}=m_{2 }$ and $m_{2} /m_{p} = integer \geq 1$. }  This is guaranteed when making invariants by contracting indices of Cartesian tensors.

\subsection{STAR's Measurements}  

\label{sec:star}

Some time ago Kharzeev and Pisarski\cite{kahrz} ($KP$) proposed studying parity violation observing  two final state particle pairs, such as $\pi^+$ and $\pi^-$ with momenta $\vec k_{1}$, $ \vec k_{2}$. The $KP$ $P_{3D}$-odd statistic is $\vec z\cdot (\vec k_{1} \times \vec k_{2})$.   

On general grounds, any Lorentz invariant cross section that violates parity symmetry must be\footnote{The statement assumes uppolarized beams} linear in the 4-dimensional symbol $\epsilon_{\mu \nu \a \b}$. If no polarizations are measured there must be four (4) independent 4-momenta, which can be taken as two beam momenta $ p_1^\mu $, $p_2^\nu$, and two other detected momenta $k_1^\a$, $k_2^\b$.  The two beam directions are equivalent to an energy (timelike) direction $E^{\mu} =(E,$ 0, 0, 0) and a $z$-axis direction $p_{1}^{\mu}-p_{2}^{\mu}$.  Then \ba \epsilon_{\mu \nu \a \b} p_A^\mu p_B^\nu k_1^\a k_2^\b \ra E  (\vec P_{1}- \vec P_{2})\cdot (\vec k_{1}\times \vec k_{2}). \label{cpoddsigma}\nn \ea We call this the Ò$\epsilon$- invariantÓ; it is the Lorentz-invariant generalization of the $KP$ invariant. As mentioned in the Introduction, the role of a $z$ axis (``$z$''-symmetry) is part of {\it verifying} parity symmetry on three dimensions that does not enter the question of {\it falsifying} parity symmetry with two dimensional data.

Kinematic constraints make it impossible to find the  $\epsilon$- invariant in a $2 \ra2$ perturbative sub-processes. When only four particles are involved, conservation of momentum allows any of the four momenta in Eq. \ref{cpoddsigma} to be expressed as a linear combination of the other three. As a consequence, the  contraction with the totally antisymmetric $\epsilon^{\mu \nu \a \b}$ yields {\it zero}, a result we call ``$2 \ra2$ epsilon no-go theorem.''  The theorem stands as a barrier to simplistic $2\ra 2$ parton-level calculations that do not include initial or final state radiation, which carry away momentum, or more complicated final states. The theorem highlights a theoretical pitfall whereby Born-level calculations can yield null results, while inclusive reactions with many particles might do just the opposite.

While important, the $KP$ invariant is not unique, because any function of the momenta (or different momenta) can multiply it. Similarly, the most simple odd-$D$-parity quantity $0_{AB}^{-} =\hat x_{A} \cdot \varepsilon \cdot  \hat x_{B}$, which is equivalent to the $KP$ invariant, is not unique. It is a general fact that any invariant can be represented as a product of primitive invariants made from the smallest representation. When it comes to observables, however, the expectation of every moment-product is independent of all the rest: \ba \lan 0_{AB}^{+}0_{AB}^{-} \ran  &\neq &  \lan 0_{AB}^{+} \ran \lan0_{AB}^{-} \ran ;\nn \\ 
\lan 0_{AB}^{+}0_{CD}^{+}0_{EF}^{-} \ran  &\neq &  \lan 0_{AB}^{+} \ran \lan 0_{CD}^{+} \ran \lan O_{EF}^{-} \ran , \;\:\;\:\:\: etc.\ea Knowing every moment of all products is equivalent to knowing the full distribution. It goes without saying that a model-independent approach makes no {\it a priori} relations between moments (``cumulant analysis'').  Instead, each different moment contains information  ready to explore. 

In Ref. \cite{VoloshinParity}, Voloshin approached the question of parity-odd distributions of charged pions using the reaction-plane formalism. STAR\cite{star2010} repeated the argument.  The first step postulates pions correlated with the ``$y$'' transverse direction of angular momentum, described as  
 \ba {dN \over d \phi }= 1 
+ 2v_{1}cos(\Delta \phi) +  2v_{2}cos(2 \Delta \phi) +  2a_{1}sin(\Delta \phi) +  2v_{2}sin(2 \Delta \phi) +... \label{sigan} \ea Here $\Delta \phi =\phi-\Psi_{RP}$ introduces the assumption of one universal reaction plane angle $\Psi_{RP}$.  Refs. \cite{VoloshinParity,star2010} then turn to correlations of two particle species $a$, $b$ of the form 
 \ba \lan cos \Delta \phi_{a} cos \Delta \phi_{b} \ran - \lan sin \Delta \phi_{a} sin \Delta \phi_{b} \ran =  \lan cos( \phi_{a} + \phi_{b} -2 \Psi_{RP}) \ran \label{cordef} \\  \ra   \lan v_{1}v_{2} \ran - \lan a_{1}^{2} \ran +B_{in}-B_{out}. \nn \ea The last line is used to define $B_{in}-B_{out}$ as any terms not coming from products of Eq. \ref{sigan}. The product of two $sine$ functions led to interpreting $ \lan cos( \phi_{a} + \phi_{b} -2 \Psi_{RP}) \ran$ as a measure of charge-dependent parity violation.  This introduces the concept of Òparity fluctuationsÓ. 
 
Experimentally STAR observed surprisingly large values of $ \lan cos( \phi_{a} + \phi_{b} -2 \Psi_{RP}) \ran$. 
The STAR paper responsibly emphasizes that the correlation is actually {\it even} under parity, requiring no parity violation for its occurrence. Near the end of Ref. \cite{VoloshinParity}, its authors also carefully noted parity-conserving resonance effects that could produce the same correlation. More recently, several papers \cite{Bzdak:2009fc, Bzdak:2010fd, Pratt:2010gy,Pratt:2010zn} pointed out that azimuthal correlations stemming from transverse momentum and charge conservation can also explain the STAR data. In addition, effects of cluster particle correlations could also accommodate the data \cite{Wang:2009kd}. 

Nevertheless Ref. \cite{VoloshinParity} maintains that the quantity is a measure of strong parity-violating fluctuations, adding emphasis that certain parity-conserving simulations do not reproduce it. STAR claims that ``local parity violation cannot be significantly observed in a single event because of the statistical fluctuations in the large number of particles, which are not affected by the $P$-violating fields...'' and ``Improved theoretical calculations...are essential to understand whether or not the observed signal is due to local strong parity violation.'' 
 
We disagree. The perception that ``local parity violation cannot be significantly observed'' comes from the bias of the event-plane formalism and the focus on phenomenology of instanton approximations. Under that influence the role of Òparity fluctuationsÓ became central, despite their indirect and inconclusive nature. When exploring new physics, it is generally better not to rely on  models, and return to the basic methodology of testing symmetries.  In this sense, we find that testing strong parity \textit{violation} is at least as important as testing strong parity \textit{fluctuation}.

The model-independent meaning of $\lan cos( \phi_{a} + \phi_{b} -2 \Psi_{2}) \ran $ is given by the group classifications cited earlier, and in Table 1. There are no $0^{-}$ (parity odd quantities) involved. In coming from $2_{ab}^{+} \otimes 2_{\Psi}^{+}$ (Table 1, item 6). We find the invariant interesting, and its measurement significant,but there is no {\it model independent} basis to claim it originates in parity violation or measures parity violation in any way. Note the observable is also not available from the invariants found in single-particle distributions.  That may explain why simulation code tuned to single particle data did not reproduce its appearance.
 
The question remains why the moment seemed related to parity violation in the first place. Trig-expanding of $cos( \phi_{a} + \phi_{b} -2 \Psi_{RP})$ produces $sin(\phi_{a}-\Psi_{RP})sin(\phi_{b}-\Psi_{RP})$, among other terms. In the model of the $RP$ formalism all tensors are labeled by a single $\Psi_{RP}$, creating ambiguity. If $\Psi_{RP} \ra \Psi_{2}$ from elliptic flow, as used by STAR, then $sin(\Delta \phi_{a}) \equiv sin(\Delta \phi_{a}+\pi) =0$ vanishes identically, and the same for $b$. To save the argument one might revert to $\Psi_{RP}  \sim \Psi_{1},$ which STAR mentions examining. Under angular momentum and parity, $sin(\phi_{a}- \Psi_{1}) = \hat x_{a} \cdot \varepsilon \cdot \hat x_{\Psi}$, where $\varepsilon $ is the $2 \times 2$ antisymmetric Levi-Civita matrix. This transforms like $0^{-}$, giving an appearance of parity-odd behavior, {\it if standing alone}. Yet by Eq. \ref{cancel} it also vanishes at the single-particle, independence level where the argument is formulated. Recognizing this,
some reference to ÒfluctuationsÓ might be made, if only in a descriptive way we have not seen pinned down. Either way, the product of two such $sines$ goes like $\hat x_{a} \cdot \varepsilon \cdot \hat x_{\Psi} \, \hat x_{b} \cdot \varepsilon \cdot \hat x_{\Psi}$.  The product of two $\varepsilon$ obeys the identity \ba \varepsilon_{ij}\varepsilon_{k \ell} = \delta_{ik}\delta_{j \ell} - \delta_{i \ell}\delta_{j k}.  \nn \ea These delta functions exist from first principles, and are the ``Clebsch'' to project $\hat x_{ai}  \hat x_{bj}  \hat x_{\Psi k}  \hat x_{\Psi \ell} \ra 2_{ab}^{+} \otimes 2_{\Psi}^{+} \ra 0_{ab \Psi} $.

\section{Spin Dependent Parity Violating Observables}

\label{sec:spin}

Transverse spin effects are large in high energy reactions, creating several puzzles. Studies of transverse spin effects are currently of great interest, and may well lead to new fundamental discoveries. 

Feynman's early treatment of transverse spin in deeply inelastic scattering\cite{Feynman:1973xc} was flawed by an error in dealing with the quark mass.  A well known and related mistake was introduced by Kane, Pumplin and Repko\cite{Kane:1978nd}. By examining low-order Feynman diagrams, the group concluded that any transverse spin effect in $QCD$ must be proportional to the amplitude to flip a quark helicity, which is of order the quark mass divided by the energy $m_{a}/E$ in the diagram. The argument is wrong in general, and was never applicable in the conditions of small momentum transfers where applied, yet continues to be cited. Another argument based on Lorentz transformations was also popularized\cite{Ratcliffe:1985mp}. Under a boost along the $z$ axis a Fermion spin 4-vector's transverse projection $\vec s_{T}$ is invariant. The longitudinal projection $s_{z}$ transforms with a factor of the boost parameter $\gamma$, yielding the wrong conclusion that transverse spin effects have relative order $m_{q}/E$ once again. Meanwhile, it was known for decades\cite{Ralston:1979ys} that transverse spin effects are inherently leading order in the parton model. The selection rules of chiral symmetry {\it of hard perturbative interactions} make it necessary to measure a chirally-odd distribution using another chirally-odd distribution\cite{Cortes:1991ja}.  The conditions of high energy reactions with several momentum transfer scales are much less demanding. There is actually no feature of $QCD$ predicting any spin effects of high energy, soft strong interactions should be small. 

The STAR collaboration\cite{star:2008qb} measured large spin asymmetries of inclusive $pp \ra \pi^{0}$ at $\sqrt{s}=200$ GeV. These measurements greatly extend the reach of the pioneering Fermilab experiment E704\cite{Adams:1991cs}. The size of these asymmetries, and even larger spin-correlations reported for $\eta$ particles \cite{Heppelmann:2009cp} shows once again that spin effect have no intrinsic high energy suppression. We concentrate here on collective flow analysis, noting that multi-particle reactions should not be thought equivalent to high-precision, hard-scattering collisions designed to test electroweak theories. \footnote{Electroweak-scale parity violation in \textit{longitudinally} polarized proton collisions has been observed at 
PHOENIX \cite{Adare:2010xa, Chiu:2010zz} and STAR \cite{Aggarwal:2010vc} using the inclusive di-lepton channel at $\sqrt{s} \approx 500 GeV$}

\subsection{Transverse Spin Classifications}

Transverse spin greatly limits the number of possible new terms contributing to parity violating observables. Consider the single-spin case $p_{A }^{\uparrow} p_{B} \rightarrow X,$ where $p_{A}^{\uparrow}$ is the momentum of a polarized proton. By $LSZ$ reduction, the proton initial state $|p_{A} , \, s_{A}>$ is reduced to a Fermion interpolating field. Then the  spin 4-vector $s_A$ enters the scattering amplitude only through the terms contracted with the on-shell spinor factors \ba u(  p_A, \, s_{A}) \bar u(p_A, \, s_{A}) = (\slashed{p} + m_{p}) { 1+ \gamma^5 \slashed{s}_A  \over 2}  \nn  \ea
\textit{Thus only terms linear in $s_A$ are allowed.} This severely constrains the possible parity-conserving and parity-violating terms that depend on spin. 

Turn to azimuthal flow, mapping by projection $s_{A}^{\mu} \ra \hat s_{A} = (cos \phi_s, \,sin \phi_s)$ in the transverse plane.  Any general $n$- particle correlation with a single spin depends on some reducible tensor of the form \ba
     {\cal M}^{i...j \,k} (\hat{x}_1\, \hat{x}_2 ... \hat{x}_n; \hat{s}_A) = \hat{x}_1^i ... \hat{x}_n^j \hat{s}_A^k. \nn
\ea
Consider a single-particle spin correlation for the purpose of illustration. Parameterize it with a tensor ${\cal M}^{ij}$, 
\ba
    {\cal M}^{i \, j} = \hat{x}^i \hat{s}_A^j. \nn 
\ea
${\cal M}^{i \, j}$ decomposes into irreducible representations of $SO(2)$ as

\ba
    {\cal M}^{i \, j} &\rightarrow& {1\over 2} \left(\begin{array}{cc}  cos(\phi + \phi_s)   &  sin(\phi + \phi_s) \\  sin(\phi + \phi_s) & -  cos(\phi + \phi_s) \end{array} \right)  \nonumber \\
    &+& \frac{1}{2} cos(\phi - \phi_s) \delta^{i \, j}  
   - \frac{1}{2}sin(\phi - \phi_s) \varepsilon^{i\, j},  \label{Mtens}
\ea
where as before $\hat{x} = (cos \phi ,\, sin \phi)$.

Since it must be linear in $s_{A}$, an invariant distribution $f(\phi  \,| \, \phi_s)$ depends only on the invariant quantities associated with ${\cal M}^{i j}$, which appear in the last line of Eq. \ref{Mtens}. The distribution $f(\phi | \phi_s)$ of a $single \,particle$ then must be of the form
\ba
    f(\phi \, | \, \phi_s) =  v_0 + v_1 cos \left(\phi - \phi_s \right) + a_1 sin\left( \phi - \phi_s \right). \label{spindist}
\ea

Turn to parity. We first need to determine how three dimensional pseudo-vectors project onto two dimensions.

For every 3D vector basis set $\{\vec e_{\a} \}$ one can define a pseudo-basis in the following way: 
\[ \begin{array}{ccc}
Basis & \,\,\,\,&Pseudo-basis \\
\hline
\vec{e_1} &\,\,\,\, & \vec{e_1} \times \vec{e_2} \\
\vec{e_2} &\,\,\,\, &\vec{e_2} \times \vec{e_3} \\
\vec{e_3} &\,\,\,\,  &\vec{ e_3} \times \vec{e_1} 

\end{array} \]

Pseudo-vectors in three dimensions are linear combinations of the pseudo-basis.  Define the {\it vectorlike} 2D basis by projection $\vec{e_1} \rightarrow \hat{e_1}, \, \vec{e_2} \rightarrow \hat{e_2}$, and the {\it pseudovector-like} 2D basis by projection: 
\[ \begin{array}{ccc}
    \vec{e_1} \xrightarrow{2D} \hat{e_1} &  \,\,\,\,\,\,\,\,\,\,&\vec{e_2} \times \vec{e_3} \xrightarrow{2D} \epsilon \cdot \hat{e_2}  \\    
    \vec{e_2} \xrightarrow{2D}\hat{e_2}    &  \,\,\,\,\,\,\,\,\,\,& \vec{e_3} \times \vec{e_1} \xrightarrow{2D} \epsilon \cdot \hat{e_1}  
\end{array} \]

The pseudo-vector $\vec{s}_A$ then projects into a pure pseudo-object $\hat{s}_A$ as
\ba
        \vec{s}_A \rightarrow \hat{s}_A = s_1 \epsilon\cdot \hat{e_1} + s_2 \epsilon \cdot \hat{e_2}
\ea

Use the fact that under dihedral parity the object $\epsilon$ acquires a minus sign: $P^{T}\varepsilon \cdot P \ra -\varepsilon$. Then
\ba
    cos(\phi - \phi_s) &=& \hat{x}\cdot \hat{s}_A \xrightarrow{P_{Nk}} - cos(\phi - \phi_s) \sim 0^{-}\nn \\ 
    sin(\phi - \phi_s) &=& \hat{x}\cdot\epsilon \cdot \hat{s}_A \xrightarrow{P_{Nk}} sin(\phi - \phi_s) \sim 0^{+}. 
\ea

It is interesting this result is consistent with inversion and projection, i.e. three dimensional parity, while inversion itself is not ``transitive'' under projection. The analysis also shows a multiplicative rule for $D$-parity, with Eq. \ref{Mtens} reading $1_{x}^{\pm} \otimes 1_{s}^{\mp} \ra 2_{xs}^{-}+0_{xs}^{-}+0_{xs}^{+}$. 

The presence of a non-zero coefficient associated with $cos(\phi -\phi_s)$ would indicate parity violation in a single-particle spin distribution. Numerous other examples can be developed. By the general rule that all invariants can be reduced to products of the most primitive ones, the $D$-parity violating moments take the form $\lan 0_{AB}^{+}0_{CD}^{+}...0_{Xs}^{-} \ran$.  It is trivial to extend this to two spins. 

\subsection{Tensor Spin Correlation}

Finally, is the spin 2 irrep of ${\cal M}_{ij}$ useful? It appears so. There are many ways of constructing scalar invariants from spin-2 objects. Consider the tensor ${\cal T}_{AB}^{i \,j}$ for correlations of azimuthal directions of two final state particles (say $\pi^+$ and $\pi^-$, Table 1, Item 1, bottom). Combine this with the spin-2 components of ${\cal M}_{ij}$, Eq. \ref{Mtens}. There is one invariant we can construct out of the two spin 2 objects:
\ba
    2_{AB} \otimes 2_{ C s} \ra tr({\cal T}_{A B} \cdot {\cal M}_{C  s} ) = cos(\phi_1 + \phi_2 -  \phi_3 - \phi_s).
\ea

Despite the appearance of a $cosine$, this quantity is {\it odd} under $D$-parity. It is a parity-violating spin-dependent observable that could be measured by RHIC spin. 

Notice that it is impossible to construct an observable such as Item 6 in Table 1 from a single spin tensor correlation. The only possible combination is
\be
	 2_{AB} \otimes 2_{ s s} \ra tr({\cal T}_{A B} \cdot {\cal M}_{s  s} ) = cos(\phi_1 + \phi_2 - 2\phi_s), \nn 
\ee
but the tensor $2_{s s}$ is ruled out by the fact that only terms linear in $s_A$ are allowed. Other possible combinations can be determined using Item 5 in Table \ref{tab:Table1}, by replacing $C$ with either $A$ or $B$. A short calculation shows that the resulting observables are identical to items which can already be obtained from $1 \otimes 1$ representations already listed in the table. 

\subsection{Longitudinal Spin Classifications}
All our azimuthal observables can be extended to take into account longitudinal spin correlations. The general rule that spin must enter the distributions through a linear term only still holds, once again limiting the number of possible options. As before, the longitudinal spin parity violating observables can be constructed as
\be
	0_{AB...N}^{+}0_{XY...Z s}^-, \nn 
\ee	
where $A$ through $Z$ can be any particles in the process. 
 For instance, we can take any $D$-parity {\it even} observable involving $N$ final state particles $d^{+}(\phi_1, \phi_2 .... \phi_N)$, and multiply it with the helicity of the initial state $h(p_A, s_A)$ to get a $D$-parity {\it odd} quantity:
\be
	0_{1 2 ... N}^+  0_{A s}^-  =  d^{+}(\phi_1, \phi_2, \phi_N) h(p_A, s_A) \nn 
\ee

\section{Summary and Conclusions} 

High energy parity violation is an exciting prospect. Until signals are observed in some form or other, and certainly when signals are observed and afterwards, it will be important to use an unbiased analysis framework that is free from model-based  limitations.  

There are good reasons to seek high energy parity violation. When $QCD$ is viewed within the modern framework of gauge-covariant derivative expansions, the parity conservation of the low energy gauge sector appears to be a kinematic artifact.  That is because the ordinary $QCD$ Lagrangian cannot violate parity without violating gauge invariance. It is only by abandoning the most basic issues of boundary conditions in field theoriess that the $\theta$-vacuum, $ \theta \epsilon_{\mu \nu \a \b}tr(F^{\mu \nu}F^{\a \b})$ term gets resuscitated for theoretical discussion.  This is not always explained in the rush to construct motivation for the incredibly important task of testing parity symmetry. Since higher derivative effective actions do not have the same kinematic features, we
find it very natural to look for parity violation  at high energies, where effects are not necessarily small.   Still, parity violation, like gold, will be where it is found, and nothing here favors one model over another. 

By systematizing the subtle facts of parity in two dimensions, we were led to the dihedral group, and from that to classification of statistics under the orthogonal group.  A devious flaw of ordinary notation previously allowed distinctly different $\Psi_{m}$ to be identified with one another, concealing that each represents a distinct tensor, and each different $\Psi_{m} \equiv \Psi_{m}/(2 \pi m)$.  Attention to orthogonal group transformation properties makes this clear.  In developing a model-independent description all $\Psi_{m}$ are distinct {\it a priori}, and all moments are distinct {\it a priori}, with no conditions from ``cumulants.''    

Different $\Psi_{m}$ have been wrongly identified in the past as artifacts of statistical fluctuations.  Thus the fascinating correlations of quantities defined with different $\Psi_{m}$ remain for the most part unexplored.  We have suggested a new concept of {\it event shape sorting}.  It is the natural generalization of event-by-event statistics.  Instead of imposing a model that all events are basically elliptical relative to one ``unobservable reaction plane $\Psi_{RP}$'', we suggest letting the data inform us of event shape categories.  An interesting example concerns sorting events into right- and left-handed classes, which we illustrated by numerical simulation. 

The analysis tools are statistically robust because they involve nothing but moments of distributions.  A huge amount of data is available from current colliders, suggesting that many independent studies can be conducted We anticipate that the classifications and analysis tools can be applied equally to low or high multiplicity events at the Tevatron, $RHIC$ or $RHIC spin$, and the $LHC$.

\medskip

 \textbf{Acknowledgments:} We thank Steve Heppelmann, Steve Sanders, KC Kong and Mat McCaskey for helpful discussions. Research supported in part under DOE Grant Number DE-FG02-04ER14308.

\bibliographystyle{apsrev}

\bibliography{Pviolation_revised}

\end{document}